\documentstyle[twocolumn,aps,prl]{revtex}

\catcode`\@=11

\def\maketitle2{\par 
\begingroup
\let\cite\@bylinecite
\def\thefootnote{\fnsymbol{footnote}}%
\twocolumn[\@maketitle2\vskip2pc]%
\thispagestyle{plain}\@thanks
\endgroup
\def\thefootnote{\arabic{footnote}}%
\setcounter{footnote}{0}%
\let\maketitle2\relax \let\@maketitle2\relax
\let\@thanks\relax \let\@authoraddress\relax \let\@title\relax
\let\@date\relax \let\thanks\relax \let\@abstract\relax
\let\@pacs\relax}

\def\abstract#1{\gdef\@abstract{{\par 
\bgroup
\ifdim\prevdepth=-1000pt \prevdepth0pt\fi
\hsize\columnwidth
\dimen0=-\prevdepth \advance\dimen0 by17.5pt \nointerlineskip
\small\vrule width 0pt height\dimen0 \relax}{~~}#1\egroup}}

\def\pacs#1{\gdef\@pacs{{\par 
\bgroup
\hsize\columnwidth \parindent0pt
\ifdim\prevdepth=-1000pt \prevdepth0pt\fi
\dimen0=-\prevdepth \advance\dimen0 by20pt\nointerlineskip
\egroup} PACS numbers:~#1}}

\def\@maketitle2{
\@preprint
\@title
\ifdim\prevdepth=-1000pt \prevdepth0pt\fi
\@authoraddress
\@date
\begin{list}{}{\leftmargin=0.10753\textwidth \rightmargin=\leftmargin
\itemsep=1pc\partopsep=-1pc}
\item\@abstract
\item\@pacs
\end{list}
}

\catcode`\@=12

\begin{document}
\draft

\title{Nonextensive Thermostatistics and the $H$-Theorem}

\author{J. A. S. Lima$^{1}$, R. Silva$^{1}$, and A. R. Plastino$^{2,3}$}

\address{$^1$Departamento de
F\'\i sica Te\'orica e Experimental,
Universidade Federal do Rio Grande do Norte, \\
59072-970, Natal-RN, Brazil
\\$^2$Facultad de Ciencias
Astronomicas y Geofisicas,
Universidad Nacional de La Plata, \\
C.C. 727, 1900 La Plata, Argentina
\\$^3$Departament de Fisica,
Universidat de les Illes Balears,
07071 Palma de Mallorca, Spain}

\date{\today}

\abstract
{\small{The kinetic foundations of Tsallis'
nonextensive thermostatistics are
investigated through Boltzmann's
transport equation approach. Our analysis
follows from a nonextensive generalization
of the ``molecular chaos hypothesis".
For $q>0$, the $q$-transport equation
satisfies an $H$-theorem based on Tsallis
entropy. It is also proved that the
collisional equilibrium is given by Tsallis'
$q$-nonextensive velocity distribution.
}}


\maketitle2
\narrowtext

In 1988 Tsallis proposed a striking  
generalization of the Boltzmann-Gibbs entropy functional
given by \cite{T88}, 
\begin{equation}\label{Sq}
S_q \, = \, - k\sum_i p_i^{q} \, \ln_q p_i \quad,
\end{equation}
where $k$ is Boltzmann's constant, $p_i$ is the probability 
of the $i$-th microstate, the parameter $q$ is any 
real number, and the $q$-logarithmic function is defined 
as \cite{T99,B98}
\begin{equation}\label{eq:224}
\ln_q f\, =\, (1-q)^{-1}(f^{1-q}-1), \,\,\,\,\,\,\,\,\, (f>0).
\end{equation}
For future reference it is convenient to introduce 
the $q$-exponential function $e_q(f)$, which is defined by 
\begin{equation} \label{eq:123}
e_q(f)=[1+(1-q)f]^{1/1- q}\quad,
\end{equation}
if $1+(1-q)f>0$ and by $e_q(f)=0$ otherwise.
These $q$-functions satisfy, for $f,g>0$, the identities (see \cite{B98} for a thorough discussion)
\begin{eqnarray}\label{Inv}
e_q(\ln_q f) &=& f\quad, \cr
\ln_q f \, + \, \ln_q g \, &=& \, 
\ln_q fg \, + \, (q-1) (\ln_q f)(\ln_q g)\quad.
\end{eqnarray}
When $q\rightarrow 1$ all the above expressions reproduce those 
verified by the usual elementary functions and Tsallis' entropy 
reduces to the standard logarithmic one, 
namely: $S_1 =-k\sum_i p_i\ln p_i$. The most distinctive trait
of $S_q$ is its pseudoadditivity. Given two systems $A$ and $B$ 
independent in the sense of factorizability of the (joint) microstate
probabilities, the Tsallis entropy of the composite system $A \oplus B$
verifies $S_q(A \oplus B) = S_{q}(A) + S_{q}(B) + (1-q)S_q(A)S_q(B)$. 
Hence, $|1-q|$ quantifies the lack of extensivity of $S_q$.

The $q$-thermostatistics associated with $S_q$ \cite{T88,T99} 
is nowadays being hailed as the possible basis of a theoretical 
framework
appropriate to dealwith nonextensive settings. There is 
a growing body of evidence 
suggesting that $S_q$ provides a convenient frame for the 
thermostatistical analysis of many physical systems and processes
\cite{B95,LKRQT98,B00,LRJ00,BGM99}, 
such as 
the velocity distribution of galaxy clusters \cite{LKRQT98},
Landau damping in plasmas \cite{LRJ00},
superdiffusion phenomena \cite{BGM99} 
and, more generally, 
systems exhibiting a nonextensive thermodynamic behaviour due,
for instance, to long range interactions \cite{T88,T99}. It is 
worth to stress that some of the aforementioned 
developments involve a quantitative agreement between 
experimental data and  theoretical models based on 
Tsallis' thermostatistics \cite{B95,LKRQT98,B00,LRJ00}.
For instance, it was experimentally found that pure 
electron plasmas in Penning traps relax to metaestable
states whose radial density profiles do not maximize
Boltzmann-Gibbs entropy. However, Boghosian showed 
that the observed profiles are well described by 
Tsallis' thermostatistics with $q$ close
to $1/2$ \cite{B95}. Beck's recent treatment of fully
developed turbulent flows constitutes another interesting
example \cite{B00}. Based on Tsallis formalism
with $q\ne 1$, Beck calculated
the probability density functions of velocity
differences depending on distance and Reynolds number, as
well as the concomitant scalling exponents, finding good 
agreement with turbulence experiments.

A large portion of the experimental evidence supporting 
Tsallis proposal involves the non-Maxwellian  (power-law) 
velocity $q$-distribution associated with Tsallis'
generalized canonical ensemble approach to
the classical $N$-body problem. 
This equilibrium $q$-distribution may be 
derived through a simple nonextensive generalization of the Maxwell 
ansatz \cite{nois}. Alternatively, it can be obtained maximizing 
Tsallis entropy under the constraints imposed by normalization 
and the energy mean value \cite{C99}, 
a procedure closely related to Jaynes information theory 
formulation of statistical mechanics \cite{J57,BS93}. 
So far, most theoretical studies on Tsallis thermostatistics  
have been developed on the basis of
the maximum entropy principle \cite{T88,TLSM95,RML98,ST99}.
As widely known, this approach to statistical ensembles was, 
in the case of standard statistical mechanics, 
historically the latest one to appear. Even when Gibbs' introduced his
ensemble approach, the kinetic foundations of statistical 
mechanics were already well developed. 
On the light of this, it is not unreasonable to expect that a systematic 
exploration of the kinetic aspects of Tsallis' 
thermostatistics may be crucial to illuminate its foundations as well as to 
achieve a better understanding of its physical applications. 

In this framework, the aim of this letter is to obtain the
equilibrium velocity $q$-distribution from a
slight generalization of the kinetic
Boltzmann $H$-theorem. The whole
argument follows simply by modifying the
molecular chaos hypothesis, as
originally advanced by Boltzmann, and 
generalizing the local entropy
expression in according to  Tsallis
proposal.

The statistical content
of Boltzmann's kinetic theory relies on two
main ingredients\cite{S93,T79}. The first one
is a specific functional form for the local entropy, 
which is expressed by Boltmann's logarithmic measure 

\begin{equation} \label{Boltzmann1}
H[f] \, = \, -k \, \int \, f(\vec x, \vec v, t) \,
\ln f(\vec x, \vec v, t) \, d^3v  \quad.
\end{equation}

\noindent
The second one is the celebrated hypothesis of
molecular chaos (``Stosszahlansatz"), which is 
tantamount to assume the factorizability of the
joint distribution associated with two
colliding molecules

\begin{equation} \label{Boltzmann2}
f(\vec x_1, \vec v_1, \vec x_2, \vec v_2, t) \, = \,
f(\vec x_1, \vec v_1, t) \, f(\vec x_2, \vec v_2, t) \quad.
\end{equation}

\noindent
These two statistical assumptions are inextricably intertwined. 
Therefore, if one adopts a generalized nonextensive entropic measure, a consistent generalization of the ``Stosszahlansatz" hypothesis should also be implemented.  Different choices for the collision term in the kinetic 
equation (which, in turn, is determined by the ``Stosszahlansatz")
lead to different forms for the entropic
functional exhibiting a time derivative with
definite sign. As a consequence, the form of the entropic
functional behaving monotonically with time (and
consequently admitting an $H$-theorem) depends upon the
form of the collisional term appearing in the kinetic equation.

Historically, the basic assumption
(\ref{Boltzmann2}) (sometimes referred to 
as ``Maxwell's ansatz") has generated a
lot of controversy. The fundamental role played
by this hypothesis  was first realized  by S.H.
Burbury \cite{B1894} in 1894. The
precise characterization of the
conditions of its applicability
has been an important conceptual
problem of theoretical physics 
ever since \cite{B83}. The physical
meaning of equation (\ref{Boltzmann2})
is that colliding molecules are uncorrelated.
The irreversible
behaviour of Boltzmann's
equation can be traced back to
this assumption. It is clearly a time asymmetric
hypothesis, since molecules assumed to be 
uncorrelated {\it before} a collision
certainly become correlated
{\it after} the collision has taken
place \cite{Z92}.

Although very plausible, and endowed with
an intuitively clear statistical meaning, Boltzmann's
particular expression (\ref{Boltzmann2}) for
 the molecular chaos hypothesis can not be
deduced from first principles. By no means it is an
inescapable consequence of classical mechanics.
Boltzmann himself accepted that
the hypothesis of molecular chaos
was needed in order to obtain
irreversibility. Further, He also admitted that the hypothesis
may not always be valid for real
gases, especially at high densities \cite{B83,B64}.

In what follows we introduce a consistent
generalization of this hypothesis in accordance
with Tsallis' nonextensive formalism. We remark
that equation (\ref{Boltzmann2}) implies that
the logarithm of the joint distribution
$f(\vec x_1, \vec v_1, \vec x_2, \vec v_2, t)$
is equal to the sum of two terms, each one
involving only the marginal distribution
associated with one of
the colliding molecules. Our generalized
hypothesis is to assume that  {\it a power} of
the joint distribution ({\it instead
of the logarithm}) is equal to
the sum of two terms, each one
depending on just one of the colliding
molecules. By recourse to
the $q$-generalization of the logarithm
function, this
condition can be formulated in a way
that recovers the standard
hypothesis of molecular chaos as a limit case.

Let us now consider a spatially
homogeneous gas of $N$ hard-sphere
particles of mass $m$ and diameter $s$, under the
action of an external force $\vec{F}$,
and enclosed in a volume $V$. The state of a
non-relativistic gas is kinetically characterized by
the one-particle distribution
function $f(\vec{x},\vec{v},t)$. The quantity
$f(\vec{x},\vec{v},t)d^{3}x d^{3}v$
gives, at each time t, the number of
particles in the volume element
$d^{3}xd^{3}v$ around the particle
position $\vec{x}$ and velocity
$\vec{v}$. In principle, this
distribution function
verifies the $q$-nonextensive
Boltzmann equation
\begin{equation}
\label{Beq}
{\partial f\over\partial t} +
\vec{v}\cdot{\partial f\over\partial
\vec{x}}+{1\over m}\vec{F}\cdot{\partial f\over
\partial\vec{v}}=C_q(f) \quad,
\end{equation}
where $C_q$ denotes the $q$-collisional term.
The left-hand-side  of (\ref{Beq}) is
just the total time
derivative of the distribution function. Hence,
nonextensivity effects can be incorporated only
through the collisional term.
Naturally, $C_q(f)$ may be
calculated in accordance with the
laws of elastic collisions. Its
specific structure must lead to
the standard result in the limit $q\rightarrow 1$.
We also make the following assumptions:
(i) Only binary collisions
occur in the gas; 
(ii) $C_q(f)$ is a
local function of the slow varying
distribution function;
 (iii) $C_q(f)$
is consistent with the energy,
momentum and particle number
conservation laws.

 Our main goal is now to show that 
the generalized collisional term
$C_q(f)$ leads to a nonnegative expression 
for the time derivative of the $q$-entropy, 
and that it does not vanish unless the 
distribution function assumes the equilibrium form 
associated with $q$-Maxwellian gas \cite{nois}.
Now, following standard lines we define
\begin{equation} \label{eq:2.11}
C_q(f)={s^2\over 2}\int |\vec V \cdot \vec e| R_q d\omega d^{3}v_1\quad,
\end{equation}
where $d^{3}v_1$ stands for the volume element in
 velocity space,  $\vec V= \vec v_1 - \vec v$ is the relative velocity
before collision, $\vec{e}$ denotes an arbitrary unit
vector, and $d\omega$ is an
elementary solid angle such that
$s^2d\omega$ is the area of
the ``collision cylinder"
(for details on the collision's geometry see
Refs.\cite{S93,T79}). The quantity
$R_q(f,f')$ is a difference of two
correlation functions (just before and
after collision), which are
assumed to satisfy a $q$-generalized
form of the molecular chaos hypothesis.
In the present $q$-nonextensive
scenario we will assume that

\begin{eqnarray} \label{eq:2.12}
\lefteqn{R_q(f,f') = e_q({f'}^{q-1}\ln_q f'+{f'}_1^{q-1}\ln_q
f'_1)}
\nonumber \\
&& -e_q(f^{q-1}\ln_q f+
f_1^{q-1}\ln_q f_1)\quad,
\end{eqnarray}
where primes refer to the distribution
function after collision. When $q \rightarrow 1$ 
equation (\ref{eq:2.12}) reduces to Boltzmann's 
molecular chaos hypothesis 
\begin{equation} \label{eq:2.15}
\lim_{q \rightarrow 1}{R_q}=R=f'{f'}_1-ff_1\quad.
\end{equation}

For the local entropy we adopt Tsallis expression,
\begin{equation} \label{eq:1}
H_q=-k\int f^q\ln_q f d^{3}v\quad ,
\end{equation}
which reduces to the standard Boltzmann measure 
(\ref{Boltzmann1}) for $q=1$. Now, we first take
the partial time derivative of the
above expression
\begin{equation} \label{eq:14}
{\partial H_q\over\partial t}
=-k\int[qf^{q-1}\ln_q f +1]{\partial
f\over\partial t}d^{3}v\quad.
\end{equation}
As one may check, by inserting the generalized
Boltzmann equation (\ref{Beq}) into
(\ref{eq:14}), and using ({\ref{eq:2.11}), 
expression (\ref{eq:14})
can be rewritten as a balance equation
\begin{eqnarray*} \label{eq:2.19}
{\partial H_q\over\partial t} +
\nabla\cdot\vec{S_q}=G_q({\vec r}, t)\quad,
\end{eqnarray*}
where the $q$-entropy flux 
vector $\vec{S_q}$ associated with $H_q$ is  defined by
\begin{equation}\label{eq:2.18}
\vec{S_q}=-k\int\vec{v}f^q\ln_q fd^{3}v\quad,
\end{equation}
and the source term $G_q$ reads
\begin{eqnarray*} \label{eq:2.20}
G_q=-{k s^2\over 2}\int |\vec V \cdot \vec e|
(1+ q f^{q-1}\ln_q f)R_q d\omega
d^{3}v_1d^{3}v\quad.
\end{eqnarray*}
In order to rewrite $G_q$ in a more symmetrical form some elementary 
operations must be done in the above expression. Following standard 
lines \cite{S93}, we first notice that interchanging $\vec{v}$ and 
$\vec{v_1}$ does not affect the value of the integral. This happens
because the magnitude of the relative velocity vector and the scattering 
cross section are invariants. Similarly, the value of $G_q$ is not altered 
if we integrate with respect to the variables $\vec{v'}$ and $\vec{v_1'}$ 
(we recall that $d^{3}v_1d^{3}v=d^{3}v'_1d^{3}v'$). Note that this step 
requires the change of sign of $R_q$ (inverse collision). Implementing these
operations and symmetrizing the resulting expression, one may show that 
the source term can be written as 
\begin{eqnarray}
\lefteqn{G_q({\vec r},t)=-{ks^2\over 8}
\int |\vec V \cdot \vec e|
(qf_1^{q-1}\ln_q f_1+qf^{q-1}\ln_q f}\nonumber \\
  &&  -q{f'}_1^{q-1}
     \ln_q {f'}_1-q{f'}^{q-1}\ln_q f')
     R_q d\omega d^{3}v_1d^{3}v\quad.
\end{eqnarray}
Making now the transformation 
$f^{q-1}\ln_{q}f = \ln_{q^*}f$, 
where $q^{*}=2-q$, and 
rearranging terms we find
\begin{eqnarray*}\label{eq:2.26a}
\lefteqn{G_q={ks^2q\over 8}
\int |\vec V \cdot \vec e|(\ln_{q^*} {f'}+\ln_{q^*}
{f'}_1-\ln_{q^*} f-\ln_{q^*} f_1)}\\
 && [e_q(\ln_{q^*}f'+\ln_{q^*}
{f'}_1)-e_q(\ln_{q^*}f+\ln_{q^*}f_1)]
d\omega d^{3}v_1d^{3}v\quad.
\end{eqnarray*}
Note that the integrand in the above equation is never negative,
because $(\ln_{q^*} f'+\ln_{q^*}
{f}'_1-\ln_{q^*} f-\ln_{q^*} f_1)$ and
$[e_q(\ln_{q^*}f'+\ln_{q^*}{f'}_1)
-e_q(\ln_{q^*}f+\ln_{q^*}f_1)]$ always
have  the same signs. Therefore, for
positive values of $q$, we obtain 
the $H_q$-theorem
\begin{equation} \label{eq:2.28}
{\partial H_q\over\partial t} +
\nabla\cdot\vec{S_q}=G_q({\vec r},t)\ge 0.
\end{equation}
 This inequality states that the $q$-entropy 
source must be positive
or zero, thereby furnishing a kinetic
argument for the second law of thermodynamics
in the framework of Tsallis' nonextensive
formalism. However, our argument does not constitute a
{\it kinetic proof} of the second law. As happens with the standard 
Boltzmann equation, our generalization can not be
obtained only from the Hamiltonian equations of
motion. Specific statistical assumptions are also needed.

When $q<0$ the entropy of a given volume element decreases 
with time. Consequently, it seems that within the present
context, and according to the second law of thermodynamics, 
the parameter $q$ should be restricted to positive values \cite{GP}. 
Notice also that the entropy does not change with time if
$q=0$. Similar results were previously obtained using the master 
equation and the relaxation time approximation \cite{M92}. 
Naturally, Tsallis' $q$-parameter may be further 
restricted by other physical requirement, such as a finite total 
number of particles. In point of fact, appropriate normalization  
of Tsallis' distribution requires a $q$-parameter greater 
than 1/3 \cite{LSS00}.

To complete the proof, we now show that Tsallis' equilibrium 
$q$-distribution  \cite{nois} is a natural consequence of the $H_q$-theorem. 
As happens in the canonical $H$-theorem, $G_q=0$ must be a necessary and 
sufficient condition for equilibrium. Since the integrand appearing in the 
expression of $G_q$ cannot be negative, this occur if and only if
\begin{equation} \label{eq:2.29}
\ln_{q^*}f'+\ln_{q^*}{f'}_1 =
\ln_{q^*}f+\ln_{q^*}f_1 \quad.
\end{equation}
Therefore, the above sum of $q$-logarithms  remains
constant during a collision: it is a summational invariant.
Only the particles total mass, energy, and momentum 
behave like that \cite{S93,T79}. Consequently, we must have
\begin{equation} \label{eq:2.32}
\ln_{q*} f = a_o+\vec{a_1} \cdot \vec{v}+a_2\vec{v} \cdot \vec{v}\quad,
\end{equation}
where $a_o$ and $a_2$  are constants and
$\vec{a_1}$ is an arbitrary constant vector. By
introducing
the barycentric velocity, $\vec{u}$, we may
rewrite (\ref{eq:2.32}) as

\begin{equation} \label{eq:2.33}
\ln_{q*} f=\alpha-\gamma^{*}{|\vec{v}-\vec{u}|}^2 \quad,
\end{equation}

\noindent
with a different set of constants. Taking
$A_{q^*}=e_{q^*}(\alpha)$ and defining 
$\gamma \, = \, \frac{\gamma^{*}}{(1-q^{*})\alpha}$, 
we obtain a generalized Maxwell's distribution

\begin{equation} \label{eq:2.34a}
f_0(\vec{v})=A_{q^*}[1-(1-q^*)
\gamma|\vec{v}-\vec{u}|^2]^{1/1-q^*}\quad,
\end{equation}

where  $A_{q^*}$, $\gamma$ and $\vec{u}$
may be functions of the temperature. The above expression is the 
general form of the $q$-Maxwellian
distribution function\cite{nois}.

Summing up, we have discussed a $q$-generalization
of Boltzmann's kinetic equation along the lines of
Tsallis nonextensive thermostatistics.
Our main results followed from a slightly modified 
version of the {\it statistical hypotheses} underlying 
Boltzmann's approach, incorporating
(i) the nonextensivity property, explicitly introduced trough
a new functional form for the local entropy, and 
(ii) a nonfactorizable expression for the molecular chaos hypothesis. 
Both ingredients were shown to be consistent with
the standard laws of (microscopic) dynamics. They
reduce to the familiar Boltzmann assumptions
in the extensive limit $q\rightarrow 1$. 
The usual statistical hypothesis
of completely uncorrelated colliding molecules seems to be
too restrictive. It is conceivable that correlations may be 
relevant within some scenarios. Here we have provided
a simple type of correlations that makes sense within Tsallis'
nonextensive thermostatistics. Other possibilities, also leading
to Tsallis' distribution (\ref{eq:2.34a}), are obtained if one 
replaces the function 
$e_q(x)$ in (\ref{eq:2.12}) by other positive, 
increasing function $F_q(x)$ such
that $\lim_{q\rightarrow 1} F_q(x) = \exp(x)$. Naturally, 
these $q$-generalizations of
the molecular chaos hypothesis do not settle the profound conceptual
issues raised by Boltzmann's ``Stozssahlansatz". What we are advocating 
is that Boltzmann's statistical assumptions do not encompass all
the possibilities allowed by the general principles of mechanics.

The study of chaotic, low dimensional dissipative 
dynamical systems has suggested a deep connection 
between Tsallis formalism and multifractals (see \cite{T99}
and references therein). It would be interesting to explore 
if this relationship also holds for Hamiltonian systems of 
large dimensionality and if it does, whether there is 
any connection with the $H_q$-theorem. 

Finally, we stress that the solutions of 
the generalized Boltzmann equation (\ref{Beq})
verify the $H_q$-theorem only if $q>0$. 
In that case $H_q$ is an increasing function of time and the 
time dependent solutions of (\ref{Beq})   
evolve  irreversibly towards Tsallis'  equilibrium
distribution (\ref{eq:2.34a}).  These results can be 
extended to include nonuniform systems as well as
more general interparticle interactions. 

\noindent
{\bf Acknowledgements:}
This work was supported by the Pronex/FINEP (No. 41.96.0908.00), CNPQ and 
CAPES (Brazil), and by CONICET (Argentina).

\end{document}